\documentclass[11pt,preprint]{aastex}

\begin{document}

\title{{\it FUSE} Observations of Molecular Hydrogen in Translucent
Interstellar Clouds: II. The Line of Sight Toward HD 110432}

%% Use \author, \affil, and the \and command to format
%% author and affiliation information.
%% Note that \email has replaced the old \authoremail command
%% from AASTeX v4.0. You can use \email to mark an email address
%% anywhere in the paper, not just in the front matter.
%% As in the title, you can use \\ to force line breaks.

\author{Brian L. Rachford\altaffilmark{1},
Theodore P. Snow\altaffilmark{1},
Jason Tumlinson\altaffilmark{1},
J. Michael Shull\altaffilmark{1},
E. Roueff\altaffilmark{2},
M. Andre\altaffilmark{3},
J.-M. Desert\altaffilmark{4},
R. Ferlet\altaffilmark{4},
A. Vidal-Madjar\altaffilmark{4},
and Donald G. York\altaffilmark{5},
}

\altaffiltext{1}{CASA, University of Colorado, 389 UCB, Boulder, CO
80309}
\altaffiltext{2}{DAEC, Observatoire de Meudon, F-92195 Meudon, France}
\altaffiltext{3}{Department of Physics and Astronomy, The Johns Hopkins
University, 3400 N. Charles St., Baltimore, MD 21218}
\altaffiltext{4}{Institut d'Astrophysique de Paris, CNRS, 98bis, Blvd.
Arago, F-75014 Paris, France}
\altaffiltext{5}{Astronomy and Astrophysics Center, University of Chicago,
5640 S. Ellis Ave., Chicago, IL 60637}

\begin{abstract}
We report the second study from the {\it FUSE} survey of molecular
hydrogen in translucent clouds, for the line of sight toward HD 110432.
This star lies beyond the Coalsack dark nebula, and with $E(B-V)$ = 0.40,
and $A_V$ = 1.32 this line of sight bridges the gap between less
extinguished diffuse cloud lines of sight with $A_V$ $\sim$ 1, such as
$\zeta$ Oph, and the translucent clouds with $A_V$ $\gtrsim$ 2
such as HD 73882.
Through profile fitting and a curve-of-growth analysis, we have
derived rotational populations for H$_2$ for $J$ = 0--7.  The line
of sight has a total molecular hydrogen column density, log N(H$_2$)
= 20.68 $\pm$ 0.05 cm$^{-2}$, nearly identical to that toward $\zeta$ Oph, but
a factor of three less than HD 73882.  The ratio of N($J$=1) to
N($J$=0) yields a kinetic temperature T$_{\rm kin}$ = 63 $\pm$ 7 K,
similar to other lines of sight with $A_V$ $\gtrsim$ 1.  The high-$J$
lines show considerable excitation above this temperature.  However,
the high-$J$ lines cannot be well-fit to a single excitation
temperature, and the even-$J$ lines of para-hydrogen appear slightly
enhanced relative to the odd-$J$ lines of ortho-hydrogen.  The high-$J$
excitation is similar to that toward $\zeta$ Oph, but much
smaller than that toward HD 73882.  Chemical modeling indicates
that the physical conditions in the cloud(s) are very similar
to those in the cloud(s) toward $\zeta$ Oph.
An analysis of {\it IUE} spectra of the Lyman-$\alpha$
line gives log N(\ion{H}{1}) = 20.85 $\pm$ 0.15 cm$^{-2}$.  Combined
with N(H$_2$), we derive a hydrogen molecular fraction, $f_{\rm H2}$ =
0.58 $\pm$ 0.12, statistically identical to that found
for the lines of sight toward $\zeta$ Oph and HD 73882.  From
the {\it FUSE} data, and a curve-of-growth analysis using the same
component structure as H$_2$ we find log N(HD) = 15.2$^{+0.7}_{-0.4}$
cm$^{-2}$, between the
values found for $\zeta$ Oph and HD 73882.  Profile fitting suggests
smaller $b$-values for HD than for H$_2$ and a value log N(HD) =
16.0$^{+0.2}_{-0.3}$ cm$^{-2}$.  From {\it FUSE} and {\it IUE} data
we derive log N(CO) $\approx$ 14.3 cm$^{-2}$ assuming the same
component structure as CH, or log N(CO) $\approx$ 14.8 cm$^{-2}$ if
all the observed CO is co-spatial with the strongest CH component.
From the combined measurements of hydrogen and carbon-containing
molecules, the line of sight toward HD 110432 appears
quite similar to the diffuse cloud line of sight toward $\zeta$ Oph,
and quite dissimilar to the translucent cloud line of sight toward HD
73882.  Upcoming {\it FUSE} observations will further explore the
transition between diffuse and translucent clouds.

\end{abstract}


\section{Introduction}
The wavelength coverage and sensitivity of the {\it Far-Ultraviolet
Spectroscopic Explorer} ({\it FUSE}) opens new
windows for the study of interstellar material toward fainter
stars with more extinction than have been studied previously
in the far-UV (Moos et al.\ 2000).  In particular, this wavelength
range gives access to the numerous transitions of molecular
hydrogen (H$_2$) and deuterated molecular hydrogen (HD) below
1130 \AA\ that cannot be studied with other existing instruments. 
In addition, the sensitivity is much greater than previous instruments
capable of observations at similar wavelengths such as {\it Copernicus}
and {\it ORFEUS}.

In Snow et al.\ (2000; hereafter Paper I) we presented an analysis
of H$_2$ for HD 73882, the first translucent cloud line of sight
observed with {\it FUSE}.  For this line of sight, with $A_V$ = 2.44, we
found an H$_2$ total column density, log N(H$_2$) = 21.08, about
3 times as great as that found for lines of sight with $A_V$ $\sim$ 1.
The hydrogen molecular fraction, $f_{\rm H2}$, for HD 73882 was comparable to
the largest values found for diffuse cloud lines of sight.
In a companion paper, Ferlet et al.\ (2000) presented an analysis
of HD for the same line of sight, finding log N(HD) = 16.1, but with
large uncertainty.  Again, this value is considerably larger than
that found for $\zeta$ Oph-like lines of sight.

The {\it FUSE} translucent cloud survey program involves 36 lines of
sight with $A_V$ = 1--5, and is described in more detail in Paper I.
In the current paper we present our second analysis of H$_2$ and
HD from this program, for the line of sight toward HD 110432.  With
$A_V$ = 1.32, this line of sight represents the transition between
diffuse clouds like those along the line of sight toward $\zeta$ Oph,
and the translucent cloud lines of sight with larger extinction such
as HD 73882.  Given that some of the material that contributes to
the extinction may not be co-spatial with the molecular gas, a
larger $A_V$ does not guarantee that the H$_2$ and HD measurements
are sampling a different environment.  Good-quality data exist for HD
110432 for the entire wavelength range covered by {\it FUSE} (912--1187
\AA ), and this is the most complete study to date of H$_2$ for a line
of sight with significant reddening.

In \S\ 2 we discuss previous measurements along this line of sight.
In \S\ 3 we present a detailed description of our analysis procedures.
In \S\ 4, 5, and 6 we present, discuss, and summarize our results,
respectively.

\section{The line of sight toward HD 110432}
HD 110432 lies beyond the southern Coalsack at a distance of 300
$\pm$ 50 pc ({\it Hipparcos}; ESA 1997).  At $V$ = 5.24,
HD 110432 is one of the brightest stars lying behind substantial
interstellar material with substantial UV flux not observed
by {\it Copernicus}.

\subsection{Spatial distribution of IS material along the line of
sight}
While considering the material between us and the star, we
must carefully consider the star itself.  HD 110432 is a Be star
with a somewhat uncertain spectral type.  Literature values 
derived from photometry and spectroscopy range in subtype from
O9 to B2, and in luminosity class from III to V.  The ionized
circumstellar material that gives rise to the emission-line
behavior also contaminates determinations of interstellar color
excess and extinction, fundamental parameters for the comparison
of lines of sight.

Codina et al. (1984) derived $E(B-V)$ = 0.40 and a spectral type
of B0.5 IIIe from {\it IUE} spectra.  However, they also derive a
distance of 430 $\pm$ 60 pc, somewhat larger than the {\it Hipparcos}
distance.  Seidensticker (1989) performed a comprehensive analysis
of stars in the Coalsack region, deriving $E(B-V)$ = 0.52 for
HD 110432, a photometric spectral type of O9 Ve, and a distance of
400 $\pm$ 120 pc.  However, the color excess was anomalously high
for its distance relative to other stars in the immediate region.
Most interestingly, Dachs, Engels, \& Kiehling (1988) derived $E(B-V)$ =
0.49, but attributed 0.09 magnitudes of this color excess to the
circumstellar material.  The remaining color excess is identical to
that of Codina et al. (1984) which was unaffected by the circumstellar
material at such short wavelengths, and should reliably characterize
the material producing the interstellar absorption lines.  At the
same time the {\it total} color excess agrees well with the value
from Seidensticker (1989), whose method would have included the
circumstellar component.  We thus adopt $E(B-V)$ = 0.40 to represent
the interstellar material containing the H$_2$ and HD.

The UV photometric extinction curve for HD 110432 was classified
as ``peculiar'' by Meyer \& Savage (1981).  The normalized
extinction, $E(\lambda - V)/E(B-V)$, for this line of sight is
much smaller than the Galactic average, particularly for the shortest
wavelength measurements at 1550 and 1800 \AA .  This result
suggests a much above-average value of total-to-selective extinction,
$R_V$.  However, as we have already noted, the Be nature of this
star must be considered.  Meyer \& Savage (1981) used $E(B-V)$ =
0.51, but with our preferred interstellar value of $E(B-V)$ = 0.40,
the interstellar extinction curve appears much closer to the
Galactic average.  Part of this correction may be offset by adjustment
of $V$ in the calculation of $E(\lambda - V)$, but even this effect
is fractionally smaller for a fixed $\Delta V$ at the shorter
wavelengths due to the larger color excesses at those wavelengths.

The wavelength of maximum polarization, $\lambda_{\rm max}$, has been
shown to be linearly correlated with $R_V$ (Serkowski, Mathewson, \&
Ford 1975).  In addition, while the observed polarization towards Be
stars may be influenced by local polarization of circumstellar material,
this effect is small in the face of substantial interstellar material,
and should not affect $\lambda_{\rm max}$ (Whittet \& van Breda 1978).
Serkowski et al.\ (1975) found $\lambda_{\rm max}$ = 0.59 $\mu$m,
corresponding to $R_V$ = 3.3, slightly larger than the Galactic
average of 3.1.  Combined with our adopted color
excess, we obtain $A_V$ = 1.32, consistent with the value from
Seidensticker (1989) minus the effects of the circumstellar
material.

The absorption along this line of sight is dominated by the
Coalsack, a prominent dark nebula centered near the same RA and
Dec as HD 110432.
The distance to the nebula has been derived by photometric analysis
of stars in the field by many authors.  The value of 174 $\pm$ 14 pc
from the early photographic study of Rodgers (1960) has held up well.
Franco (1989) derived 180 $\pm$ 26 pc from photoelectric data.
Seidensticker \& Schmidt-Kaler (1989) considered more stars
and found evidence for two sub-clouds at distances of 188 $\pm$ 4
and 243 $\pm$ 14 pc with the latter cloud providing somewhat
greater extinction than the former.  The morphology of the extinction
map agrees well
with the CO emission map of Nyman et al.\ (1989).  No matter what
the true distribution of material may be, at 300 pc HD 110432 lies
behind all of this material with very little extinction between
the Coalsack and the star.  In addition, foreground material
represents a small fraction of total absorption along the line of
sight ($A_V$ $\sim$ 0.1 mag; Seidensticker \& Schmidt-Kaler 1989).

\subsection{Cloud velocity structure}
An accurate assessment of the H$_2$ velocity
structure is crucial for the interpretation of the high-$J$ lines,
and the resolution of the {\it FUSE} spectra is generally insufficient
for this purpose.  Since individual components typically have velocity
spread parameters, $b$ $\sim$ 1--2 km s$^{-1}$, resolution of comparable
magnitude is preferred.  Fortunately, HD 110432 has been observed at
ultra high resolution ($\Delta v$ = 0.3 km s$^{-1}$), at optical
wavelengths (Crawford 1995).  These observations provided velocity
structures for CH, CH$^{+}$, \ion{Ca}{2}, and \ion{K}{1}.  Of most
relevance to H$_2$ are CH and \ion{K}{1}.  

The CH observations reveal a simple velocity structure with
just two components separated by 4.0 km s$^{-1}$.  The \ion{K}{1}
profile is somewhat more complex, with five components, dominated
by two strong components separated by 1.3 km s$^{-1}$.  The combined
profile of the two closely-spaced strong \ion{K}{1} components plus
one of the weak components rather closely matches the overall shape
of the strong CH component.  Further, the profile of the remaining two
weak \ion{K}{1} components spans the same velocity range as the
weaker CH component, albeit this latter component did not appear
to be double.  Since the overall line profiles are generally
similar, the ``effective'' $b$-values for the two profiles are also
similar, 2.4 km s$^{-1}$ for CH and 2.1 km s$^{-1}$ for \ion{K}{1}.
Given the low molecular hydrogen kinetic temperature we derive (63 K;
\S\ 4), the velocity dispersion is dominated by turbulent motion and
not thermal motion.  Thus, scaling the CH $b$-values to H$_2$ according
to mass only increases the effective $b$-values by about 10\%, and
these scaled values were used to generate the adopted component structure
given in Table 1.  While we use the CH structure to model the H$_2$
structure,
the \ion{K}{1} structure would give similar results.  In \S\ 4 we
describe two tests which support this choice of velocity structure
for H$_2$.

Other species have been observed in absorption at resolutions of
$\Delta v$ $\sim$ 3 km s$^{-1}$, and Table 2 gives a summary of the
column densities of previously observed molecules and atoms.  As
van Dishoeck \& Black (1989) first pointed out, the column densities
of molecules such as C$_2$, CH, and CN are more typical of diffuse
cloud lines of sight such as $\zeta$ Oph, than translucent clouds.
The value of N(CO) from Codina et al.\ (1984) is very uncertain,
and in \S\ 5 we derive a more precise value based on additional
{\it IUE} spectra unavailable to those authors, and revised
oscillator strengths.

Millimeter-wave CO measurements with a 43$\arcsec$ beam show little
if any emission at the velocities of the previously mentioned absorption
features (Gredel et al.\ 1994), but do show a strong component at
$v_{\rm LSR}$ = 5.1 km s$^{-1}$ ($v_{\rm helio}$ = $-$2.5 km s$^{-1}$),
about 5 km s$^{-1}$ blue-shifted from the bluest absorption components.
As pointed out by Crawford (1991), the absorption line data support
the conclusion of Nyman et al. (1989) that the CO emission
at this velocity near HD 110432 represents background material.

\section{{\it FUSE} Observations and Data Analysis}
HD 110432 was observed by {\it FUSE} on 2000 Apr 4--5 for a total
of 3631 seconds spread across 5 short integrations.  Due to the
relatively large UV-flux, the data were recorded in spectral image
mode (Moos et al.\ 2000).
In this mode, integration times are kept short to prevent detector
and spacecraft motion from degrading the spectral resolution.
Excellent data were recorded on all 8 detector segments, with
stellar signal recorded essentially down to the Lyman limit as
Figure 1 indicates.
The data were processed with version 1.6.9 of the CALFUSE pipeline,
but we have used a later, improved version of the wavelength solution.
We have co-added the 5 integrations for each segment, propagating
the pixel-by-pixel uncertainties through the co-additions.
Given the differences in the wavelength scales and line-spread
functions of the different detector segments, we have not
combined data from different segments even though the wavelength
ranges overlap.  We emphasize that while the spectrum depicted in
Figure 1 was formed by combining all detector segments, this
spectrum is primarily useful for display purposes and is not
suitable for precise measurements.

With the full wavelength coverage of {\it FUSE} available for
this target, a wealth of information on H$_2$ is available.
The spectrum is dominated by the heavily damped and blended
$J$ = 0 and $J$ = 1 lines from the Lyman series (B--X) vibrational
bands from (0,0) through (18,0), and the Werner series (C--X)
vibrational bands from (0,0) through (4,0).  At shorter wavelengths,
the strongly damped Lyman series \ion{H}{1} lines become progressively
more important.  The H$_2$ lines for $J$ = 2 also show damping
wings, while the progressively weaker $J$ = 3 through $J$ = 7
line profiles are dominated by the apparently Gaussian line spread
function.  Given the various H$_2$ line profiles encountered, 
we use multiple techniques to derive column densities, and
we describe those techniques in detail as they will be used
in the analysis of all additional targets in this program.

At total H$_2$ column densities greater than $\sim$ 10$^{20}$
cm$^{-2}$ the overlapping profiles from adjacent bands conspire
to create a spectrum where the true continuum level is never
reached.  Figure 2 shows a normalized model spectrum for H$_2$
corresponding to the $J$ = 0 through $J$ = 5 column densities we
derive for HD 110432.  The effect of the overlapping bands
is smallest for the longest
wavelength vibrational bands due to the combination of having fewer
contributing bands on the red side, and smaller oscillator strengths
for these bands.  Shortward
of 1010 \AA , the effect is magnified by overlapping Lyman and
Werner bands.  Most of our program stars have even larger
column densities with correspondingly greater effect.  An important
consideration then is that the continuum can not be located
{\it a priori}.

Thus, to determine the column densities for $J$ = 0 and $J$ = 1,
we must perform profile fits to as many vibrational bands as is
practical, with the continuum being a fit parameter.  However,
there are several complicating factors.  First, the saturated cores
of $J$ = 0 and $J$ = 1 blends provide no leverage in the continuum
determination.  Second, the H$_2$ lines with $J$ $\geq$ 2, as
well as lines from other species must be included in the fit.
Third, some H$_2$ profiles are contaminated by stellar lines,
as well as idiosyncrasies in the large-scale response of the detectors.
Fourth, the effects of $J$ = 0 and $J$ = 1 line wings from adjacent
bands cannot be ignored.  Finally, blends between the Lyman and
Werner series H$_2$ lines, and blends of H$_2$ and \ion{H}{1} lines
limit the number of H$_2$ bands that can be fitted with the highest
accuracy.

To overcome the first factor, we must include a significant
wavelength range in the fits, typically about 10 \AA\ per band.
For the second factor, we include all H$_2$ lines up to $J$ = 5,
as well as identifiable lines of other atomic and molecular
species.  The weaker lines carry little weight in the overall
fits, and they are not always modeled as accurately as if they
were treated individually, but do allow for accurate continuum
placement.  To alleviate the other continuum problems, we removed
the obvious stellar lines by dividing by an appropriate
Gaussian profile.  Also, we only fit one band at a time such that
we can model the local continuum with a low-order polynomial.
To overcome the fourth factor, we explicitly include the effects
of all the $J$ = 0 and $J$ = 1 lines within 30 \AA\ of the
region we are fitting.  As for the selection of H$_2$ bands to
fit, the Lyman bands from (0,0) through (4,0) are clean of
Werner bands and \ion{H}{1} lines.  Each of these bands occurs
in 2, 3, or 4 detector segments.

In the particular case of HD 110432, we were unable to obtain
good fits for the (0,0) band, presumably due to stellar
contamination in the lower portions of the wings, or at the narrow
``bump'' between the R(0) and R(1) lines.  Without resorting to
unverifiable guesses as to the nature of contamination,
or a full non-LTE modeling of the stellar spectrum, we could not
use this band for this line of sight.  However, in other
program stars with different (generally earlier) spectral types,
this effect may not be important.  Even without the (0,0) band,
we still have a total of 13 profiles from which to derive the
low-$J$ column densities.  Figure 3 shows a sample fit of the
(2,0) band covering the 1072--1082 \AA\ region.

For $J$ $\geq$ 2, the lines are weak enough that one can measure
individual equivalent widths by fitting Gaussian profiles to each
line, or resolved blends of lines.  These measurements are relative
to the observed background level; i.e. the $J$=0--1 line wings.  We
were able to measure at
least one individual H$_2$ line of $J$ $\geq$ 2 from each
Lyman-series vibrational band from (0,0) through (18,0), and from
each Werner-series vibrational band from (0,0) through (4,0).  In
total, we obtained 249 equivalent width measurements for 99 H$_2$
transitions.  In 22 cases, we were able to measure a transition
in the maximum possible 4 detector segments, and for all but 15
transitions we measured the line in at least 2 segments, allowing
us to check the data for self-consistency.

The lines of HD behave much like the high-$J$ lines of H$_2$.
Typically, the equivalent widths of the $J$ = 0 lines of HD are
similar to those of the $J$ = 4 and $J$ = 5 lines of H$_2$.  The
$J$ = 1 lines of HD are considerably weaker, and conclusive detection
of such lines in our {\it FUSE} spectra will generally be difficult.
For HD 110432, we obtained 24 equivalent width measurements for 8 $J$ =
0 lines, and none for $J$ = 1.

We determine an error estimate for each line measurement based on the
formal uncertainty of the Gaussian fit parameters, combined in quadrature
with a formal estimate of the uncertainty in continuum placement.
In combining measurements of the same line from different detector
segments, we must be careful how we treat systematic differences.
A simple weighted mean of the individual measurements will
provide encouragingly small uncertainties when several measurements
are combined, but if the actual values cover a large range, this
reported uncertainty is not useful.  However, the standard deviation
is not useful considering the small number of measurements.
The procedure we have adopted takes the
largest of the three following numbers: one-half of the total range
of measurements for a particular line (a proxy for the standard
deviation), the error we derive from a weighted mean,
and 10 per cent of the weighted mean itself.  The latter limit
provides a conservative estimate of any systematic errors that
we may not have considered.  There are six pairs of detector
segments whose wavelength coverage overlap.  In comparing equivalent
widths for the same line in these pairs of segments, we find
excellent agreement with an exact one-to-one relationship.
The only possible exception is for the SiC 1A and SiC 1B segments
where the strongest lines were consistently measured to be 10--15\%
stronger than in other segments.  However, this only involves a
half-dozen lines, mostly from $J$ = 2 where we used profile
fitting for the final column densities, as discussed below.

From the final equivalent width measurements, and the assumed velocity
structure, we construct a multi-component curve of growth.  Using
the uncertainties of the individual lines, we perform a least-squares
fit for each $J$ to determine N($J$) for H$_2$ and HD.

The $J$ = 2 lines of H$_2$ represent a special case.  They are strong
enough to have noticeable damping wings, yet the surrounding continuum
is frequently not clean enough to allow accurate fitting of those wings
with an individual damped Voigt profile.  The Gaussian profile fitting,
used in the measurement of the weaker lines, underestimates the
equivalent widths of the strongest $J$ = 2 lines by as much as 10--15\%
for HD 110432, which
corresponds to a 0.1--0.2 dex underestimate of the column density.
However, the profiles of these damped lines are fitted well along with
the $J$ = 0 and $J$ = 1 profiles, which gives us slightly better results
than the curve of growth analysis for this particular line of sight.

The confusion between the H$_2$ bands and the \ion{H}{1} lines
makes a determination of N(\ion{H}{1}) somewhat difficult using
{\it FUSE} data.  However, this method may eventually produce
accurate column densities, especially for cases where N(\ion{H}{1})
$>>$ N(H$_{2}$); i.e. where the hydrogen molecular fraction is small.
In the case of HD 110432, we have used {\it IUE} spectra of
Lyman-$\alpha$ to determine N(\ion{H}{1}).  We co-added all 13
available SWP spectra from the {\it IUE} archive to improve the
S/N.  We used the technique of Bohlin (1975) to determine the
best column density by matching a pure damping profile to the
observed line wings.  The primary uncertainty in this method
beyond the noise in the data is the presence of stellar lines
in the wings of Lyman-$\alpha$, particularly the blue wing.
As we discuss in \S\ 4, the actual component structure can only
have a small effect on the derived N(\ion{H}{1}).

\section{Results}
We begin with the results of the band-by-band profile fits.
Table 3 summarizes these results.  Fits of the same band in
different detector segments show exceptional agreement.  
However, band-to-band differences exist.  In all segments, the
(3,0) band yields larger column densities for $J$ = 0.  The mean
of the (3,0) values is about 0.15 dex larger than the mean from
the other bands.  The $J$ = 2 column densities are also slightly
larger than those from the other bands, but the $J$ = 1 columns
are not significantly larger.  A visual inspection of the (3,0)
band fits does not reveal an clear explanation for the differences,
although one or more stellar lines may be the culprit.  We have
included these values in the final averages given in Tables 3 and
4 although it may be appropriate to exclude this band in future
studies.  The reported uncertainties for $J$ = 0--2 are
simply the sample standard deviation of the 13 individual values.
If we exclude the (3,0) results, the uncertainty is cut in half
for $J$ = 0, while the average value decreases by 0.04 dex.

Figure 4 shows the 2-component curve of growth along with our
averaged measurements of 99 H$_2$ transitions and 8 HD transitions.
Since the uncertainties
in N($J$) for H$_2$ lines of $J$ $\geq$ 3 and the HD lines are
affected by the shape of the curve of growth, we have given two-sided
1-$\sigma$ error bars to these values in Table 4.  For HD 110432,
most of the error bars do not show much asymmetry.  We emphasize that
the uncertainties in these values of N($J$) only apply to the assumption
that we have a perfect description of the component structure.
Changes in the component structure can produce changes
in column densities when the lines lie on or near the flat
portion of the curve of growth.  For instance, if we had used
the \ion{K}{1} structure, our derived column densities for $J$ = 3--6
would be increased by 0.10, 0.20, 0.25, and 0.10 dex, respectively.
While these changes are modest, if we have seriously misjudged
the H$_2$ velocity structure, the changes could be much greater.

With this point in mind, we performed two additional tests of
the validity of our assumed velocity structure.  First, given
the H$_2$ equivalent width data, we fitted
a single-component curve of growth, allowing both the column
densities and the $b$-value to vary.  The best-fit $b$-value, and
in turn the column densities, were consistent with the ``effective''
$b$-value and column densities of the two-component model.  Second,
we performed profile fits to several bands assuming
a two-component structure with velocity separation given by the CH
data, but allowing the $b$-values and relative strengths of the
components to vary.  Again, the derived component parameters were
consistent with the CH structure.  Both tests indicate that the
changes in the column densities due to possible uncertainty in the
component structure are not larger than the differences between
the CH and \ion{K}{1} structure indicated in the previous paragraph .

Our best curve-of-growth column density for $J$ = 2 is 18.55,
or 0.13 dex smaller than our profile fits, consistent with our
previous discussion of the fitting of the $J$ = 2 lines.
The column density for $J$ = 7 is based on only one very
weak line.  Also, N($J$=6) is based on just 4 weak lines
with very little spread in $\lambda f$.  The small observed
uncertainty thus may not be reliable.

Our fit of the \ion{H}{1} Lyman-$\alpha$ line gives log
N(\ion{H}{1}) = 20.85 $\pm$ 0.15 cm$^{-2}$.  We note that
if instead of assuming a pure damping profile, we use the
CH velocity structure and scale the b-values according to
mass, log N would be about 0.05 dex smaller.  However, even
a slight increase in the complexity of the \ion{H}{1} component
structure relative to CH could decrease this difference to
near zero.

Given the H$_2$ column densities (particularly for $J$ = 0--5),
and the \ion{H}{1} column density,
we can explore the physical conditions of the absorbing gas.
When combined with log N(\ion{H}{1}) = 20.85 $\pm$ 0.15 cm$^{-2}$,
our total H$_2$ column density of log N(H$_2$) = 20.68 $\pm$
0.05 cm$^{-2}$ gives a molecular fraction $f_{\rm H2}$ = 0.58
$\pm$ 0.11; i.e.\ between one-half and two-thirds of the observed
hydrogen atoms exist in H$_2$ molecules.  At the
relatively high densities assumed to exist in these H$_2$ cloud(s),
the ratio of N($J$=1) to N($J$=0) is considered an indicator of the
kinetic temperature of the gas, and we derive $T_{\rm kin}$ = 63
$\pm$ 6 K.

The excitation diagram in Figure 5 shows the enhanced populations
for $J \geq 2$ relative to the kinetic temperature of the gas.
This non-thermal excitation is usually interpreted as pumping
by UV photons, followed by cascading transitions down through
the various rotational states (Black \& Dalgarno 1973).  The high-$J$
lines in many lines of sight can be fitted by a single excitation
temperature, i.e. a straight line in the excitation plot (Spitzer \&
Cochran 1973).  For HD 110432 such a fit for $J$ = 2--5 seriously
overestimates N($J$=3), and in general the even-$J$ lines (para-hydrogen)
appear slightly enhanced relative to the odd-$J$ lines (ortho-hydrogen).
However, good linear relationships do appear for $J$ = 1--3 and $J$ = 3--5.
For the former levels, we derive a temperature T$_{13}$ = 110 $\pm$
5 K, and for the latter, we derive T$_{35}$ = 240 $\pm$ 40 K.  
An even less certain fit for $J$ = 3--7 corresponds to T $\approx$ 350 K.

\section{Discussion}
Several lines of sight with $A_V$ $\sim$ 1 were observed with
{\it Copernicus}, $\zeta$ Oph being the prototype.  We will present
a detailed comparison of all lines of sight with $A_V$ $\gtrsim$ 1,
including more than a dozen of our {\it FUSE} translucent cloud
program stars in a future work (B. L. Rachford et al. 2001, in
preparation).  In the current
paper, we will generally limit our comparisons to $\zeta$ Oph (Morton
1975; Savage et al.\ 1977), HD 73882 (Paper I), and HD 110432, with
only brief references to the larger survey.  Relevant data for these
lines of sight appear in Table 5.  The lines of sight have similar
total-to-selective extinction ratios, but cover a range of more than
a factor of 2 in color excess and thus visual extinction.

The kinetic temperatures of the three lines of sight are statistically
identical.  In addition, our preliminary survey of {\it FUSE}
translucent cloud program stars indicates H$_2$ kinetic temperatures
of $\sim$60 K are typical for such lines of sight.  The ratio
N(H$_2$)/$A_V$ lies within the range 3.5--5 $\times$ 10$^{20}$
cm$^{-2}$ mag$^{-1}$ for all three lines of sight, and the molecular
fractions are nearly identical as well.  However, these latter
two ratios do not hold up in the larger survey.  In particular,
we see a large range in $f_{\rm H2}$ with a possible upper limit
near 0.7.  If verified, this upper limit may provide important
information on formation and destruction mechanisms at H$_2$ column
densities near 10$^{21}$ cm$^{-2}$.

The combination of line saturation and the relatively limited
range of log $f\lambda$ produces a very uncertain curve-of-growth
value of N(HD) that falls squarely between the values for $\zeta$
Oph on the low side, and HD 73882 on the high side.  The HD/H$_2$
ratio of 3 $\times$ 10$^{-6}$ falls well below the expected value
of 2$\times$D/H that would be observed if HD represents the
primary reservoir of deuterium.  However, if HD occurs primarily
in the cores of the clouds(s) along this line of sight, the
H$_2$ component structure may not be appropriate.  Fits of the
HD lines assuming a 2-component model that allow the strengths
and $b$-values to vary, support this view.  The best fits yield
log N(HD) = 16.0$^{+0.2}_{-0.3}$ cm$^{-2}$, with b-values of 1.4
and 1.2 km s$^{-1}$ for the blue and red components, respectively,
and a much weaker blue component than for H$_{2}$.  This result
gives an overall HD/H$_2$ ratio slightly smaller than typical
values of 2$\times$D/H.  The HD/H$_2$ ratio for the red component
would be $\sim$4 $\times$ 10$^{-5}$, quite similar to 2$\times$D/H.
However, we emphasize that the uncertainty is about a factor of 2.

As depicted in Figure 6, the excitation of the high-$J$ levels of
H$_2$ is similar to, but slightly larger than that of $\zeta$ Oph;
yet is much smaller than HD 73882.
Two factors control this excitation, the UV interstellar radiation
field (ISRF) and the number density.

HD 110432 itself is an early enough star to excite H$_2$, and with
the two-cloud structure of the Coalsack proposed by Seidensticker
(1989), the star lies 60 $\pm$ 50 pc beyond the cloud at $\sim$240 pc.
The other cloud, at $\sim$190 pc, would then lie 110 $\pm$ 50 pc from
the star.  To estimate the effects of HD 110432 on the Coalsack, we
have taken fluxes at 1000 \AA\ from an appropriate Kurucz (1979) model
for a B0.5 III star (T$_{\rm eff}$ = 25000 K, log $g$ = 3.5, solar
metallicity), a stellar radius of 15R$_{\sun}$, and then compared the
resultant radiation field at various distances from the star to the
average ISRF (Mathis, Mezger, \& Panagia 1983).  Assuming no dimunition
of the stellar radiation field between the star and the cloud(s),
we find that at about 40 pc from the star the stellar field is equal
to the average ISRF\footnote{If HD 110432 is actually of spectral type
B2 Ve, this distance drops to around 10-15 pc.}.  We can then safely
assume
that HD 110432 does not contribute significantly to the radiation
field of the cloud at 190 pc, but we cannot make the same assumption
about the other cloud.  Even if the material is more uniformly
distributed between 180 and
250 pc, the same conclusion holds; the material more distant from
Earth may be affected by the radiation field from HD 110432 while the
material nearest to Earth is not.

We have searched the {\it Hipparcos} catalog for other O and B
stars in the region.  In Figure 7, we present distances and
spectral types for stars within about 6 degrees of HD 110432.
There are many stars much closer than the Coalsack, and many more
beyond the nebula, but few
at the appropriate distances to be exciting the cloud(s) along this
line of sight.  At 174 pc, HD 110737 (302.2,$-$1.8) may lie within
the nearest portion of the Coalsack, and is just 6 pc from the
line of sight to HD 110432.  However, at spectral type B9.5,
this star has many orders of magnitude less UV flux than HD 110432,
and little power to excite H$_2$.  Two early B stars, HD
103884 at (296.6,0.7) and HD 104841 at (297.6,$-$0.5), lie at the
same distance from Earth as the Coalsack, but lie 20 pc from the
line of sight, well outside the boundaries of the cloud(s).  Given
the uncertainties in the {\it Hipparcos} distances, the uncertainties
in the precise structure of the Coalsack, and the weaker UV output
as compared with HD 110432, these stars may not provide much
enhancement to the overall radiation field.
For comparison, the ISRF impinging on the cloud(s) containing H$_2$
along the line of sight to $\zeta$ Oph is thought to be no more
than twice the average field (Federman et al. 1995).

Through observations of weak, rotationally excited C$_2$ lines,
van Dishoeck \& Black (1989) estimated the number density of the
material toward HD 110432, and find an uncertain value of $n$
$\lesssim$ 200 cm$^{-3}$.  A similar analysis for $\zeta$ Oph
(van Dishoeck \& Black 1986) gives an uncertain value $n$
$\approx$ 200 cm$^{-3}$.

We have taken advantage of the amount of observational knowledge to
run chemical models of translucent clouds. The basic assumptions are
described in Le Bourlot et al. (1993). The model solves the radiative
transfer as well as the chemical molecular formation/destruction
mechanisms of a cloud containing H, C, N, O, and S species and a metal,
illuminated by the ISRF.  The
thermal balance can be solved in parallel to the chemical equilibrium.
The original model, which was devoted to study the envelopes of
molecular clouds, consists of a semi-infinite planar slab where
the radiation field is impinging on one side of the cloud. The
photodissociation of H$_2$ is taking place through absorption in the
discrete ultraviolet Lyman and Werner band transitions followed by
continuous fluorescence. The photodissociation probabilities and the
basic molecular properties of the Lyman and Werner band systems are
accurately known from previous experimental and theoretical studies
(cf. Roueff et al. 1999; Abgrall, Roueff, \& Drira 2000). Such a model
is very similar to other models of so-called Photon Dominated Regions
(PDR) (see Hollenbach \& Tielens 1999 for a recent review). It has been
recognized recently (Le Bourlot 2000) that ortho to para conversion of
molecular hydrogen may occur on grains.  Moreover, the grain temperature
depends critically on the photoelectric effect, i.e. on the actual
value of the radiation field impinging on the grains at a certain
location in the cloud . The models presented here follow the prescription
of Le Bourlot (2000) and take into account the effect of the size
distribution of the grains in the formation of H$_2$ on the grains
(Le Bourlot et al. 1995) as well as in the determination of their charge 
following Bakes \& Tielens (1994).

In order to apply the model to a diffuse cloud line of sight such as
HD 110432, we
have additionally introduced the deuterium chemistry and the detailed
mechanisms implied in the formation and destruction of HD (Le Petit,
Roueff, \& Le Bourlot 2001; in preparation) and we consider the cloud
as a finite slab of constant density submitted to the same radiation
field on both sides.
The elemental abundances are the mean interstellar values given by
Savage and Sembach (1999).  The cosmic ray ionization rate is chosen to be
5 $\times$ 10$^{-17}$ s$^{-1}$, a mean value deduced from OH observations
(Federman, Weber, \& Lambert 1996).

We assign model parameters to reproduce several observational
constraints; namely, the hydrogen molecular fraction, the
temperature-sensitive
H$_2$ column densities in $J$ = 0 and $J$ = 1, and the total visual
extinction.  The ratio between the total hydrogen column density
and $A_V$ we have derived, 1.26 $\times$ 10$^{21}$ cm$^{-2}$ mag$^{-1}$,
is slightly smaller than average.  The ISRF may be somewhat
larger than the mean standard value taken from Draine (1978) but its
actual value is not known.  The peculiar character of the extinction
curve in the ultraviolet is taken into account via the R$_V$ value of
3.3.

Within the series of models we have run, we find that the two main
constraints are fulfilled for a ratio of about 100 between the proton
density $n_H$ and the multiplicative factor $G_0$ of the ISRF
in Draine's units.  Indeed, the molecular fraction is directly
proportional to $n$/$G_0$. However, this agreement is obtained \emph{only}
when the minimum radius of the grains is larger than about 10$^{-6}$ cm.
This fact should be compared with the peculiarity of the extinction
curve: the larger than normal value of $R_V$ implies a far UV extinction
lower than the mean value, in agreement with the absence of very small grains.
This cut-off in the grain size distribution has two major effects,
one on the formation rate of H$_2$ and thus on the value of the
molecular fraction, and the other on the temperature governing the
ratio between the $J$ = 1 and $J$ = 0 column densities.
Table 6 gives the results of some typical models.

The actual value of the ISRF is not known but should not be larger than
a few; so a density of about 200 cm$^{-3}$ and a radiation field of
about twice the standard ISRF provides a reasonable compromise, in
agreement with the
density found from the study of the excitation of C$_2$.  With these
parameters, a column density of deuterated molecular hydrogen of
1.1 $\times$ 10$^{16}$ cm$^{-2}$ is obtained, in agreement with the
observations.

In contrast, none of the models reproduces the column densities of the
rotational levels of molecular hydrogen with $J$ $\geq $ 2 as shown
in Table 7.  This not surprising due to the simplified geometry assumed.
However, we feel that we address here a more fundamental failure of the
chemical model.  Indeed, the presence of the CH$^+$ molecular ion can
not be reproduced by any PDR model and another chemical component is
needed.  This puzzle has been addressed by the presence of shocks
(Pineau des For\^{e}ts et al. 1986) or by the occurrence of dissipative
structures due to interstellar turbulence (Falgarone, Pineau des For\^{e}ts,
\& Roueff 1995; Joulain et al. 1998).  The presence of an intense far
ultraviolet radiation field may also lead to an efficient formation
mechanism of CH$^+$ as suggested by Snow (1993).  However, no obvious
candidate for the production of intense ultraviolet radiation is
available in the environment of HD 110432.  Such a tiny component,
whatever its origin, will enhance the populations of excited molecular
hydrogen without modifying the populations of $J$ = 0 and $J$ = 1.
Further modeling is beyond the scope of the present paper and will
be performed subsequently.  This modeling may also reveal the
significance, if any, of the slightly enhanced column densities
of para-hydrogen relative to ortho-hydrogen.

For comparison, it appears that the density in the clouds toward
HD 73882 is larger than for the lines of sight toward $\zeta$ Oph
and HD 110432.  An
analysis of C$_2$ gives $n$ = 350$^{+300}_{-100}$ cm$^{-3}$ (Gredel,
van Dishoeck, \& Black 1993). 
In addition, the column densities of $J$ = 4 and $J$ = 5 are
the largest yet observed for any diffuse or translucent line
of sight.  This may suggest a very
strong UV field, a distinct possibility considering that HD 73882
has a spectral type of O8.5, even earlier than $\zeta$ Oph.
As noted above, the current models are unable to reproduce the
observed high-$J$ excitation, and thus an interpretation of
the extreme excitation seen toward HD 73882 will require
further work.

The column densities of carbon-containing
molecules along the line of sight toward HD 110432, such as CH,
CH$^+$, and C$_2$, more closely match those of diffuse clouds
rather than translucent clouds.  However, the CO column density
is rather poorly known and represents most of the carbon locked
up in diatomic molecules.  The millimeter wave observations of
Gredel et al. (1994) reveal little if any CO emission at the velocities
of the absorption lines of CH, etc.  Codina et al. (1984) observed
several of the CO absorption bands from the A--X series with {\it IUE}
and derived log N(CO) = 14.60 cm$^{-2}$ from a curve-of-growth analysis.
This value was quite uncertain and depended on the then poorly-known
velocity structure, and there have been subsequent revisions in the
oscillator strengths for these bands.

In principle, N(CO) can be determined from {\it FUSE} data.  Easily
visible CO bands within the {\it FUSE} wavelength range include
B--X (0,0) at 1150 \AA , C--X (0,0) at 1088 \AA , and E--X (0,0)
at 1076 \AA .  The C--X (1,0) band at 1063 \AA\ , E--X (1,0) band
at 1052 \AA , and F--X (0,0) band at 1003 \AA\ have $f$-values
large enough to be visible, but are lost in or near the cores
of strong H$_2$ $J$ = 0 and $J$ = 1 lines.

The C--X (0,0) band is contaminated by interstellar atomic lines
(\ion{Cl}{1} $\lambda$1088.06 and others).  Several weak lines lie
slightly blueward and redward of the B--X (0,0) band but reasonable
fits can be obtained in HD 110432.  In the spectrum of HD 110432, the
E--X (0,0) band lies in the blue wing of the $J$ = 0 line
of the Lyman (2,0) band of H$_2$, but is clean of narrow lines from
other species.  However, in lines of sight with much larger
H$_2$ column densities (i.e. HD 73882), this band is not visible.

We performed profile fits of the B-X (0,0) and E-X (0,0) bands including
all lines of $^{12}$C$^{16}$O and $^{13}$C$^{16}$O with $J \leq 4$
using the CH component structure.  We assumed a $^{12}$C$^{16}$O
to $^{13}$C$^{16}$O ratio of 70, a uniform rotational excitation
temperature (T$_{\rm rot}$) for all levels, and used line parameters
from Zhong, et al. (1997).  These fits yielded log N(CO) = 14.4
$\pm$ 0.4 cm$^{-2}$ and T$_{\rm rot}$ = 3.5 $\pm$ 1.3 K.
The velocity shift between the models and the spectra only differed
from the shift derived from the H$_2$ fits in this region by
$\sim$0.5 km s$^{-1}$, implying that most or all of the CO is
co-spatial with those molecules.

The CH component structure may not accurately represent the CO
component structure, similar to the situation for the HD molecule
described earlier in this section.  In fact, fits with a variable
$b$-value did suggest a smaller velocity dispersion and somewhat larger
column density.  An extreme case would be if all the CO lies within
a single velocity component.  Thus, we performed fits assuming
that the stronger CH component contains all the CO.  In this
case, we derive log N(CO) = 15.0 $\pm$ 0.5 cm$^{-2}$.

The measurements of the A-X bands given by Codina, et al. (1984)
were based on a single {\it IUE} integration.  As with our \ion{H}{1}
measurement, we have co-added the available {\it IUE} spectrum
to improve the S/N.  This co-added spectrum allows us to much
more precisely measure the equivalent widths of the A-X bands,
and these measurements, along with 1-$\sigma$ uncertaintines,
are given in Table 8.  With the 2-component
CH model, we derive log N(CO) = 14.35 $\pm$ 0.15 cm$^{-2}$.
Despite our use of smaller f-values (Morton \& Noreau 1994),
in several cases our equivalent widths were considerable smaller
than derived by Codina, et al. (1984), and thus our column
density is smaller than given by those authors.  As with our
profile fits, we also performed a curve-of-growth analysis
assuming that only the stronger CH component contains CO, and
this analysis yields log N(CO) = 14.70 $\pm$ 0.20 cm$^{-2}$.

Despite nearly identical values of N(H$_2$), N(CO) for HD 110432
is considerable smaller than for $\zeta$ Oph.  The latter value
was derived from high-resolution, high S/N spectra of the A--X
bands with {\it HST} (Lambert et al. 1994), and is a more
precise measurement.  However, even our less certain value for
HD 110432 is clearly smaller than that for $\zeta$ Oph assuming
the CH structure toward the former star.  Given the other similarities
between the two lines of sight, this difference may indicate that
indeed nearly all the CO lies within a single velocity component,
supporting the larger set of column densities we derived.

The column density of the CH radical has been shown to correlate
well with N(H$_2$) in diffuse clouds (Danks, Federman, \& Lambert 1984)
and the data point for HD 110432 falls very near the best-fit
relationship.  Interestingly, an extrapolation of this relationship
into the realm of the translucent clouds provides an excellent match
for HD 73882 as well.

\section{Summary}
We have presented the first measurements of far-UV lines of
molecular hydrogen for the line of sight toward HD 110432
($A_V$ = 1.32).  We derive a total H$_2$ column density of
log N(H$_2$) = 20.68 $\pm$ 0.05 cm$^{-2}$.  Combined with log
N(\ion{H}{1}) = 20.85 $\pm$ 0.15 cm$^{-2}$, we derive a molecular
fraction for hydrogen of $f_{\rm H2}$ = 0.58 $\pm$ 0.12.  This
value is similar to that found for the diffuse cloud line of
sight toward $\zeta$ Oph and the translucent cloud line of sight
toward HD 73882.  We find an H$_2$ kinetic temperature of
T$_{\rm kin}$ = 63 $\pm$ 6 K.  This temperature is in agreement
with those found for diffuse and translucent cloud lines of
sight with log N(H$_2$) $\gtrsim$ 20.5 cm$^{-2}$.

The excitation of the high-$J$ levels cannot be well-represented by
a single temperature model.  A two-component model, with T$_{13}$ =
110 $\pm$ 5 K, and T$_{35}$ = 240 $\pm$ 40 K can account for the
data, but does not necessarily have a physical significance.  The
even-$J$ (para-hydrogen) column densities appear slightly enhanced
relative to the odd-$J$ (ortho-hydrogen) column densities.  The column
densities in each of the rotational states are statistically
identical to those from $\zeta$ Oph.  

Preliminary chemical modeling of the line of sight toward HD 110432
indicates that physical conditions are quite similar to that
in the cloud(s) toward $\zeta$ Oph.  We derive an incident
radiation field of about twice the interstellar average, and a
density of $\sim$200 cm$^{-3}$.

The {\it FUSE} spectra allow us to investigate the diatomic
molecules HD and CO.  We obtain N(HD) = 15.2$^{+0.7}_{-0.4}$
cm$^{-2}$ from a curve-of-growth analysis of 8 $J$ = 0 lines,
yielding an
HD/H$_2$ ratio of 3 $\times$ 10$^{-6}$.  This lies between that
found for $\zeta$ Oph and HD 73882, and is much smaller than the
value implied by the D/H ratio, suggesting that most deuterium
exists outside of HD molecules in the cloud(s) toward HD 110432.
However, if HD and CO follow component structures with smaller
$b$-values than H$_2$, the column densities could be an order
of magnitude larger.  In particular, profile fits of the HD lines
suggest smaller $b$-values as compared with H$_2$, and a column
density log N(HD) = 16.0$^{+0.2}_{-0.3}$ cm$^{-2}$.  The stronger
of the two HD components then gives an HD/H$_2$ ratio very similar
to 2$\times$D/H.
Based on both a curve-of-growth analysis and profile fitting,
we find log N(CO) $\approx$ 14.3 cm$^{-2}$ if CO follows the
same component structure at CH, and log N(CO) $\approx$ 14.8 cm$^{-2}$
if all the CO is co-spatial with the stronger of the two CH components.

Overall, in terms of H$_2$, \ion{H}{1}, carbon-containing molecules,
and physical conditions, the line of sight toward HD 110432
appears very similar to that of $\zeta$ Oph, which has slightly
less visual extinction.  On the other hand, it does not closely
resemble the line of sight toward HD 73882, whose much greater
extinction and molecular abundances places it firmly within the
realm of translucent clouds.  However, preliminary measurements
of {\it FUSE} observations of other translucent cloud lines of
sight suggest that the HD 73882 line of sight may not be
representative.  Many questions regarding the nature of diffuse
and translucent clouds may be answered in upcoming studies of such
environments with {\it FUSE}.

\acknowledgments
This work is based on data obtained for the Guaranteed Time Team by the
NASA-CNES-CSA {\it FUSE} mission operated by the Johns Hopkins University.
Financial support to U.S. participants has been provided by
NASA contract NAS5-32985.  The referee provided much helpful criticism.
We also thank W. Blair for helpful comments.  This research has made
use of the SIMBAD database, operated at CDS, Strasbourg, France.

\clearpage

%% No more than seven \figcaption commands are allowed per page,
%% so if you have more than seven captions, insert a \clearpage
%% after every seventh one.

\begin{figure}
\epsscale{0.7}
\plotone{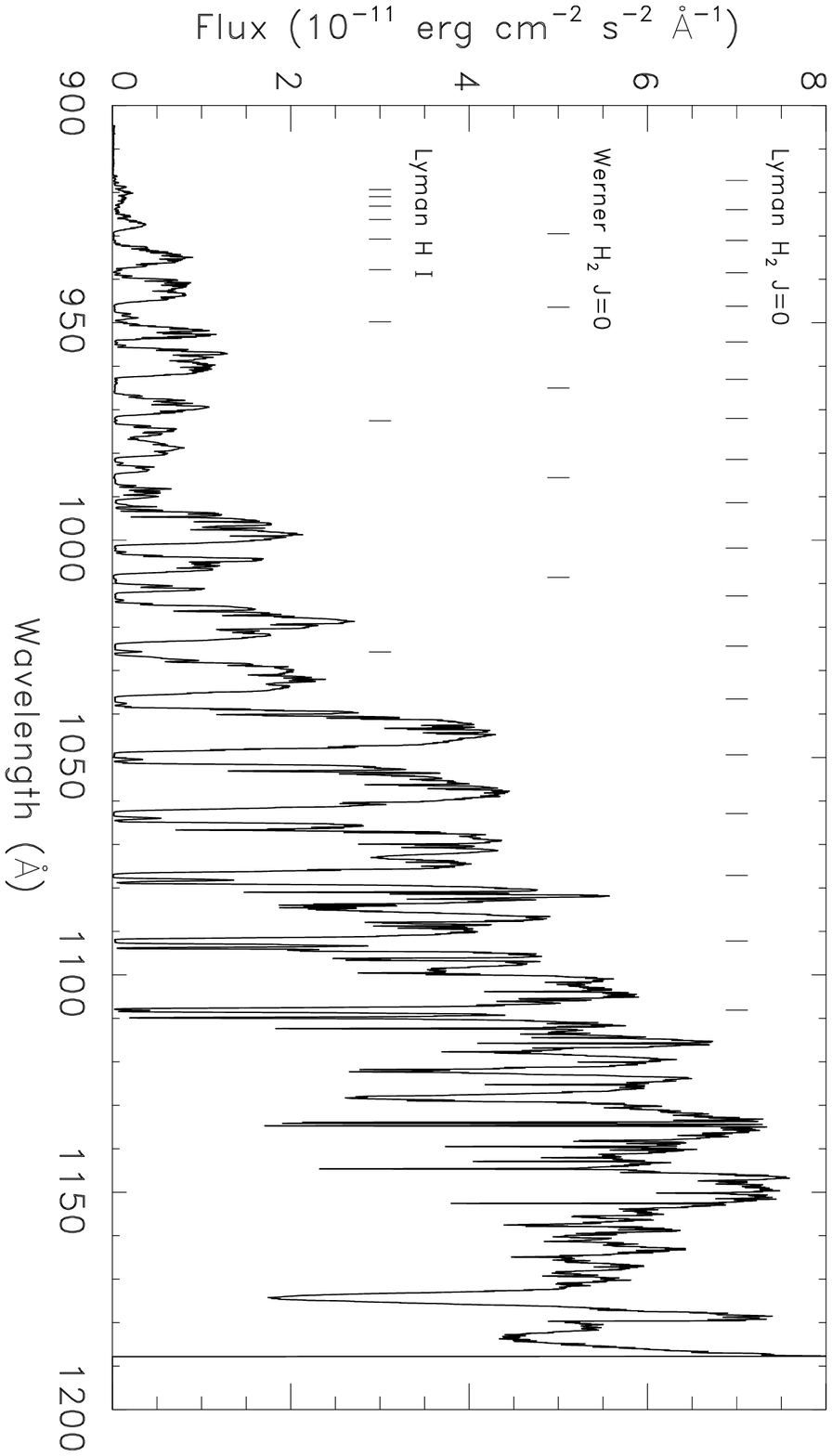}
\caption{Total merged spectrum for HD 110432 from all detector segments
and all exposures.  Lyman series H$_2$
bandheads for the (18,0) through (0,0) vibrational bands are indicated
by the top set of tickmarks.  The middle set depicts the bandheads for
the Werner series of H$_2$ for the (4,0) through (0,0) bands.  The
bottom set marks Lyman series \ion{H}{1} lines from N=10 through
$\beta$.}
\end{figure}

\begin{figure}
\epsscale{0.7}
\plotone{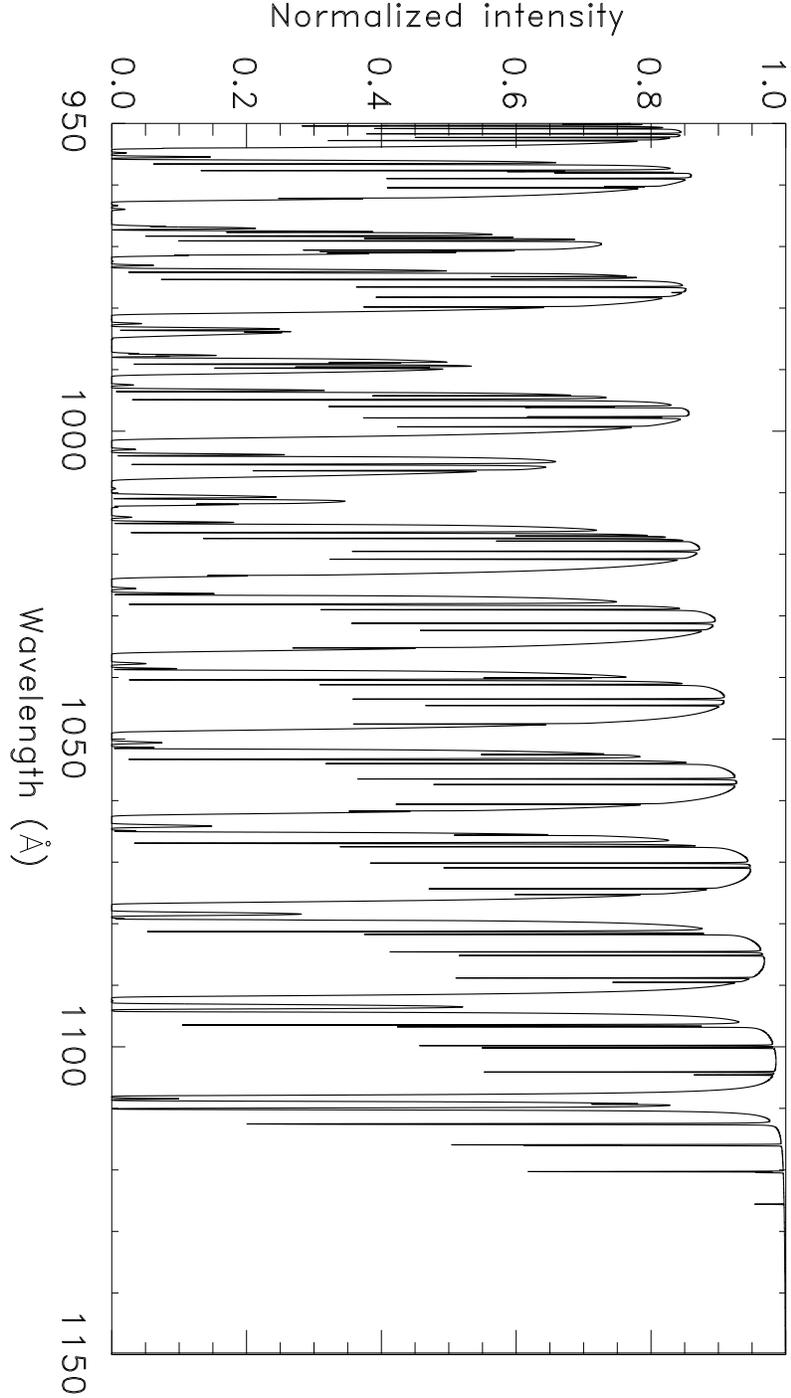}
\caption{Normalized H$_2$ model spectrum for HD 110432 based on
our derived rotational populations.  Note particularly that the
high points between vibrational bands do not reach the continuum.
We only fitted vibrational bands at the longer wavelengths where
this problem is minimized.  This model does not include \ion{H}{1}
lines which become a progressively larger problem for shorter
wavelengths, nor the stellar continuum and the effects of extinction
apparent in Figure 1.}
\end{figure}

\clearpage
\begin{figure}
\epsscale{0.7}
\plotone{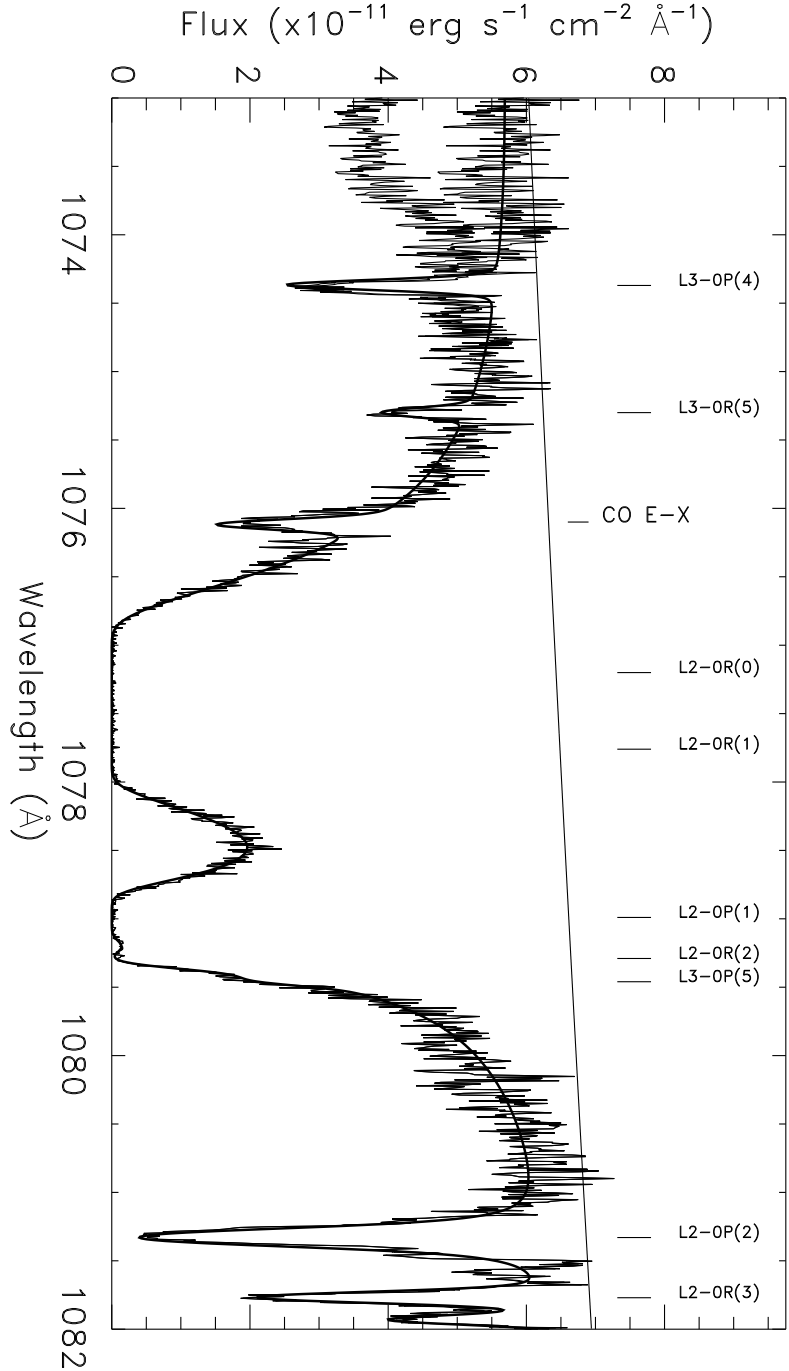}
\caption{H$_2$ profile fit for the Lyman (2,0) band from SiC 2B.
The line above the spectrum (actually slightly curved)
depicts the continuum derived in the fit.  A stellar feature
at 1073 \AA\ was removed and the plot shows both the original and
corrected spectra.  For computational simplicity, we modeled the CO
E--X (0,0) band as a single line.  While this gives a less-than-optimal
fit for the line itself, it does not affect the overall fit for the
$J$=0--2 lines.  The CO column density quoted in \S\ 5 is not based on
this simplified model.}
\end{figure}

\clearpage
\begin{figure}
\plotone{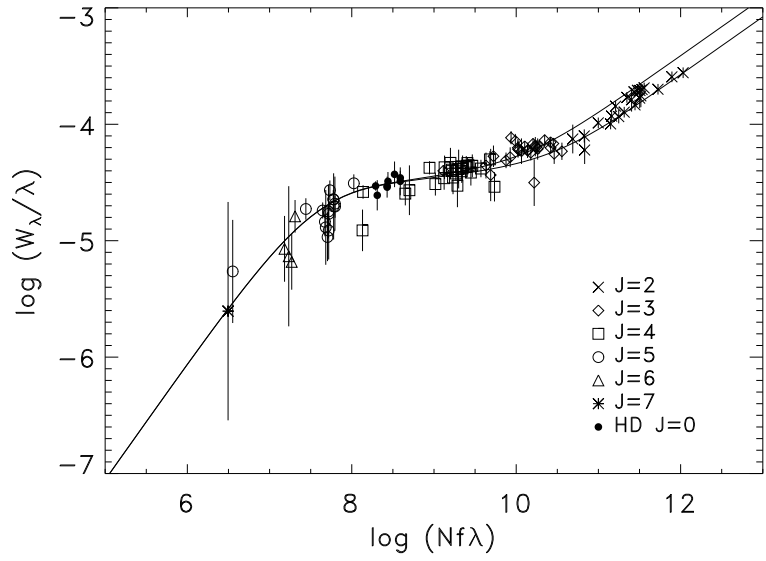}
\caption{Two-component curve of growth for HD 110432.  The two
curves represent the range in damping constants for the measured
lines.}
\end{figure}

\clearpage
\begin{figure}
\plotone{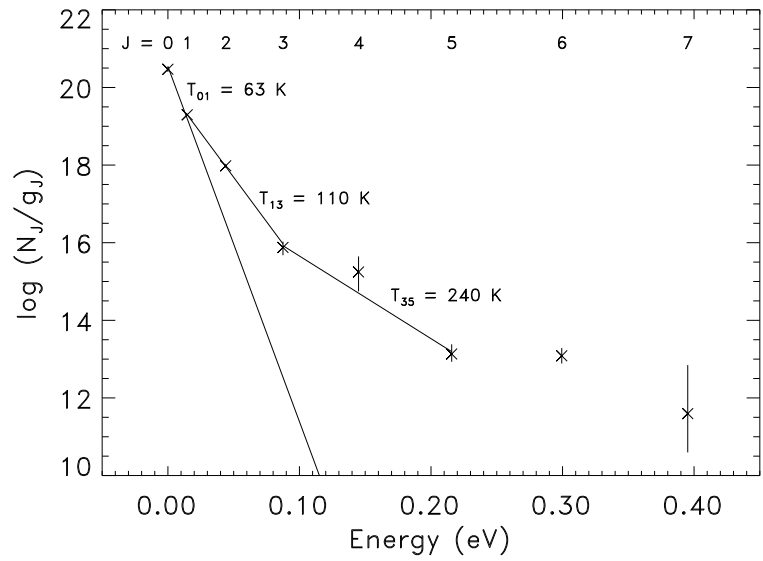}
\caption{Excitation diagram for HD 110432.  The solid lines
correspond to linear fits to $J$ = 0--1, $J$ = 1--3, and $J$
= 3--5, as described in the text.}
\end{figure}

\clearpage
\begin{figure}
\plotone{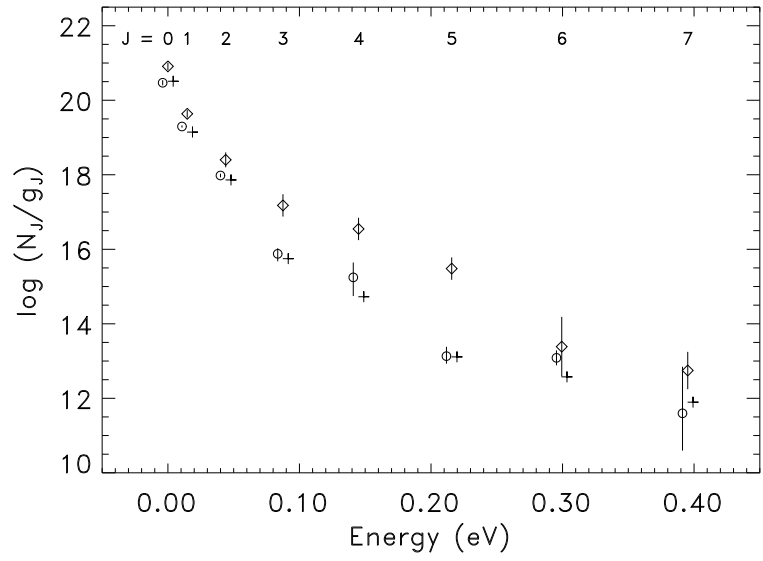}
\caption{Comparison of H$_2$ excitation in three lines of
sight.  Symbols used: HD 110432 - circles; HD 73882 -
diamonds, $\zeta$ Oph - crosses.  Error bars are not given
for $\zeta$ Oph.  Data points have been slightly offset
horizontally to separate the three datasets.}
\end{figure}

\clearpage
\begin{figure}
\plotone{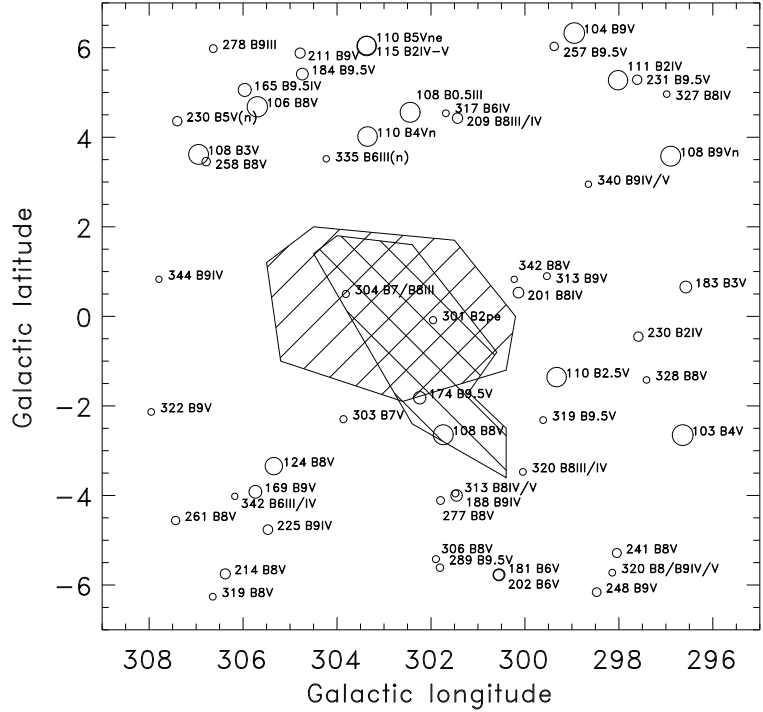}
\caption{Early-type stars near the HD 110432 line of sight.  The
symbol size is inversely proportional to the {\it Hipparcos} distance.
This distance in parsecs is given to the right of the star, along
with the spectral type.  Only stars between 100 and 350 pc were
included.  HD 110432 itself lies at the center of the
plot.  An approximate depiction of the two-cloud structure of the
Coalsack from Seidensticker \& Schmidt-Kaler (1989) is given by the
hatched areas.  The more distant cloud extends to more negative latitudes.
The scale of the plot at the distance of the nearer cloud is about
3 pc per degree.}
\end{figure}

\clearpage

\begin{deluxetable}{cccc}
\tablecaption{Adopted H$_2$ component structure}
\tablewidth{0pt}
\tablenum{1}
\tablehead{
\colhead{Comp.} & \colhead{Rel. strength} & \colhead{$b$} &
\colhead{$v_{\rm helio}$\tablenotemark{a}} \\
& & \colhead{(km s$^{-1}$)} & \colhead{(km s$^{-1}$)}
}
\startdata
1 & 0.324 & 1.8 & +\phn2.9 \\
2 & 0.676 & 1.4 & +\phn6.9 \\
\enddata
\tablenotetext{a}{Subtract 7.6 km s$^{-1}$ to obtain $v_{\rm LSR}$}
\end{deluxetable}

\begin{deluxetable}{lll}
\tablecaption{Previous absorption-line abundance data for HD 110432}
\tablewidth{0pt}
\tablenum{2}
\tablehead{
\colhead{Species} & \colhead{log N} & \colhead{Reference} \\
 & \colhead{(cm$^{-2}$)} &
}
\startdata
K I    & 11.86 & Crawford 1995 \\
Ca II  & 11.97 & Crawford 1995 \\
CH     & 13.19 & Crawford 1995 \\
CH$^+$ & 13.25 & Crawford 1995 \\
CN     & 12.08 & Gredel et al.\ 1993 \\
C$_2$  & 13.48 & van Dishoeck \& Black 1989 \\
CO     & 14.60 & Codina et al.\ 1984 \\
\enddata
\end{deluxetable}

% \clearpage

\begin{deluxetable}{cccccc}
\tablecaption{Band-by-band fits}
\tablewidth{0pt}
\tablenum{3}
\tablehead{
\colhead{Detector segment} & \colhead{Band} & \colhead{log N(0)} &
\colhead{log N(1)} & \colhead{log N(2)} \\
& & \colhead{(cm$^{-2}$)} & \colhead{(cm$^{-2}$)}
& \colhead{(cm$^{-2}$)}
}
\startdata
LiF 1A & (4,0) & 20.44 & 20.26 & 18.68 \\
LiF 2B & (4,0) & 20.39 & 20.28 & 18.67 \\
SiC 1A & (4,0) & 20.43 & 20.28 & 18.67 \\
SiC 2B & (4,0) & 20.40 & 20.27 & 18.66 \\
  Mean & (4,0) & 20.42 & 20.27 & 18.67 \\
       &      &  &  & \\
LiF 1A & (3,0) & 20.57 & 20.29 & 18.72 \\
LiF 2B & (3,0) & 20.56 & 20.27 & 18.74 \\
SiC 1A & (3,0) & 20.59 & 20.25 & 18.73 \\
SiC 2B & (3,0) & 20.55 & 20.24 & 18.71 \\
  Mean & (3,0) & 20.57 & 20.26 & 18.73 \\
       &      &  &  & \\
LiF 1A & (2,0) & 20.44 & 20.21 & 18.59 \\
SiC 1A & (2,0) & 20.40 & 20.25 & 18.71 \\
SiC 2B & (2,0) & 20.42 & 20.24 & 18.60 \\
  Mean & (2,0) & 20.42 & 20.23 & 18.63 \\
       &      &  &  & \\
LiF 2A & (1,0) & 20.49 & 20.24 & 18.68 \\
SiC 2B & (1,0) & 20.45 & 20.21 & 18.64 \\
  Mean & (1,0) & 20.47 & 20.23 & 18.66 \\
       &      &  &  & \\
  Mean &  All & 20.47 & 20.25 & 18.68 \\
\enddata
\end{deluxetable}

% \clearpage

\begin{deluxetable}{ccc}
\tablecaption{Rotational populations for H$_2$ and HD}
\tablewidth{0pt}
\tablenum{4}
\tablehead{
\colhead{Species} & \colhead{$J$} & \colhead{log N$_J$} \\
& & \colhead{(cm$^{-2}$)}
}
\startdata
H$_2$ & 0 & 20.47 $\pm$ 0.07 \\
\ldots & 1 & 20.25 $\pm$ 0.03 \\
\ldots & 2 & 18.68 $\pm$ 0.05 \\
\ldots & 3 & 17.20$^{+0.15}_{-0.20}$ \\
\ldots & 4 & 16.20$^{+0.40}_{-0.50}$ \\
\ldots & 5 & 14.65$^{+0.25}_{-0.20}$ \\
\ldots & 6 & 14.20$^{+0.20}_{-0.20}$ \\
\ldots & 7 & 13.25$^{+1.25}_{-1.00}$ \\
\ldots & Total & 20.68 $\pm$ 0.06 \\
HD\tablenotemark{a} & 0 & 15.20$^{+0.70}_{-0.40}$ \\
\enddata
\tablenotetext{a}{Curve of growth value; see text for information
on profile fits.}
\end{deluxetable}

% \clearpage

\begin{deluxetable}{cccccccc}
\tablecaption{Line of sight properties comparison}
\tablewidth{0pt}
\tablenum{5}
\tablehead{
\colhead{Line of sight} & \colhead{$E(B-V)$} & \colhead{$A_V$}
& \colhead{$R_V$} & \colhead{T$_{\rm kin}$} & \colhead{log N(H$_2$)} &
\colhead{log N(\ion{H}{1})} &
\colhead{$f_{\rm H2}$} \\
& & & & \colhead{(K)} & \colhead{(cm$^{-2}$)} & \colhead{(cm$^{-2}$)} &
}
\startdata
Zeta Oph  & 0.32 & 0.99 & 3.1 & 54 & 20.65 & 20.72 & 0.63 \\
HD 110432 & 0.40 & 1.32 & 3.3 & 63 & 20.68 & 20.85 & 0.58 \\
HD 73882  & 0.72 & 2.44 & 3.4 & 59 & 21.08 & 21.11 & 0.65 \\
\enddata
\end{deluxetable}

\begin{deluxetable}{ccccccccc}
\tablecaption{Modeled column densities}
\tablewidth{0pt}
\tablenum{6}
\tablehead{
% \multicolumn{4}{c}{Model inputs} & \multicolumn{4}{c}{log N(X)} \\
% \cline{2-5} & \cline{6-9} \\
\colhead{Model} & \colhead{$n_H$} & \colhead{$G_0$} & \colhead{$r_{min}$}
& \colhead{T} & \colhead{log N(H I)} & \colhead{log N(H$_2$)}
& \colhead{log N(HD)} & \colhead{log N(CH)} \\
& \colhead{(cm$^{-3}$)} & & \colhead{(cm)} & \colhead{(K)} &
\colhead{(cm$^{-2}$)} & \colhead{(cm$^{-2}$)} & \colhead{(cm$^{-2}$)}
& \colhead{(cm$^{-2}$)}
}
\startdata
A & 100 & 1 & 1 $\times$ 10$^{-6}$ & 70 & 20.65 & 20.74 & 16.17 & 12.89 \\
B & 100 & 1 & 3 $\times$ 10$^{-6}$ & 61 & 20.81 & 20.70 & 16.28 & 12.81 \\
C & 200 & 2 & 3 $\times$ 10$^{-6}$ & 65 & 20.79 & 20.72 & 16.04 & 12.84 \\
D & 500 & 5 & 3 $\times$ 10$^{-6}$ & 76 & 20.79 & 20.72 & 15.08 & 12.84 \\
\enddata
\end{deluxetable}

\begin{deluxetable}{cccccc}
\tablecaption{Observed vs. modeled H$_2$ high-$J$ excitation}
\tablewidth{0pt}
\tablenum{7}
\tablehead{
& \multicolumn{5}{c}{log N($J$)} \\
& \cline{1-5} 
\colhead{$J$} & \colhead{Obs.} & \colhead{Model A} & \colhead{Model B}
& \colhead{Model C} & \colhead{Model D}
}
\startdata
0 & 20.47  & 20.50   & 20.49   & 20.45   & 20.46 \\
1 & 20.25  & 20.44   & 20.28   & 20.31   & 20.42 \\
2 & 18.68  & 18.10   & 17.48   & 17.78   & 18.30 \\
3 & 17.20  & 16.57   & 15.47   & 15.76   & 16.21 \\
4 & 16.20  & 14.40   & 14.34   & 14.62   & 15.01 \\
5 & 14.65  & 13.99   & 13.89   & 14.18   & 14.60 \\
6 & 14.20  & 13.01   & 12.92   & 13.22   & 13.63 \\
7 & 13.25  & 12.95   & 12.90   & 13.18   & 13.59 \\
\enddata
\end{deluxetable}

\begin{deluxetable}{ccc}
\tablecaption{CO A-X band measurements}
\tablewidth{0pt}
\tablenum{8}
\tablehead{
\colhead{Band} & \colhead{$\lambda$} & \colhead{$W_{\lambda}$} \\
& \colhead{(\AA )} & \colhead{(m\AA )}
}
\startdata
(1-0) & 1509.76 & 76.4 $\pm$ \phn7.3 \\
(2-0) & 1477.54 & 79.0 $\pm$ 16.4    \\
(3-0) & 1447.34 & 72.0 $\pm$ 16.3    \\
(4-0) & 1419.03 & 65.5 $\pm$ 24.6    \\
(6-0) & 1367.62 & 30.6 $\pm$ 28.1    \\
(7-0) & 1344.18 & 24.2 $\pm$ 13.5    \\
\enddata
\end{deluxetable}


\begin{references}

\reference{} Abgrall, H., Roueff, E., \& Drira, I., 2000, \aaps, 141, 297
\reference{} Bakes, E. L. O. \& Tielens, A. G. G. M., 1994, \apj, 427, 822
\reference{} Black, J. H., \& Dalgarno, A. 1973, \apj, L101
\reference{} Bohlin, R. C., 1975, \apj, 200, 402
\reference{} Codina, S. J., de Freitas Pacheco, J. A., Lopes, D. F., \&
        Gilra, D. 1984, \aaps, 57, 239
\reference{} Crawford, I. A. 1991, \aap, 246, 210
\reference{} Crawford, I. A. 1995, \mnras, 227, 458
\reference{} Dachs, J., Engels, D., \& Kiehling, R. 1988, \aap, 194, 167
\reference{} Danks, A. C., Federman, S. R., \& Lambert, D. L., 1984,
        \aap, 130, 62
\reference{} Draine, B. T. 1978, \apjs, 36, 595
\reference{} ESA, 1997, The Hipparcos and Tycho Catalogues, ESA SP-1200
\reference{} Falgarone, E., Pineau des For\^ets, G., Roueff, E. 1995,
        \aap, 300, 870
\reference{} Federman, S. R., Cardelli, J. A., van Dishoeck, E. F.,
        Lambert, D. L., \& Black, J. H. 1995, \apj, 445, 325
\reference{} Federman, S. R., Weber, J., \& Lambert D. L. 1996,
        \apj, 463, 181
\reference{} Ferlet, R., et al.\ 2000, ApJ, 538, L69
\reference{} Franco, G. A. P. 1989, \aap, 215, 119
\reference{} Gredel, R., van Dishoeck, E. F., \& Black, J. H. 1993, \aap,
	269, 477
\reference{} Gredel, R., van Dishoeck, E. F., \& Black, J. H. 1994, \aap,
        285, 300
\reference{} Hollenbach, D. \& Tielens, A. G. G. M. 1999, Rev. Mod. Phys.
        71, 173
\reference{} Joulain, K., Falgarone, E., Pineau des For\^ets, G.,
        Flower D. 1998, \aap, 340, 241
\reference{} Jura, M. 1975, \apj, 197, 581
\reference{} Kurucz, R. 1979, \apjs, 40, 1
\reference{} Lambert, D. L., Sheffer, Y., Gilliland, R. L., \& Federman,
        S. R. 1994, \apj, 420, 756
\reference{} Le Bourlot, J., Pineau des For\^ets, G., Roueff, E., Flower, D.
        1993, \aap, 267, 233
\reference{} Le Bourlot, J., Pineau des For\^ets, G., Roueff, E., Flower, D.
        1995, \aap, 302, 870
\reference{} Le Bourlot, J. 2000, \aap, 360, 656
\reference{} Mathis, J. S., Mezger, P. G., \& Panagis, N. 1983, \aap,
        128, 212
\reference{} Meyer, D. M. \& Savage, B. D. 1981, \apj, 248, 545
\reference{} Moos, H. W., et al. 2000, \apj, 538, L1
\reference{} Morton, D. C. 1975, \apj, 197, 85
\reference{} Morton, D. C. \& Noreau, L. 1994, \apjs, 95, 301
\reference{} Nyman, L.-\AA ., Bronfman, L., \& Thaddeus, P. 1989, \aap,
        216, 185
\reference{} Pineau des For\^ets, G., Flower, D. R., Hartquist, T. W.,
        Dalgarno, A. 1986, \mnras, 220, 801
\reference{} Rodgers, A. W. 1960, \mnras, 120, 163
\reference{} Roueff, E., Abgrall, H., Liu, X.M., \& Shemanskii, D. 2000,
        H$_2$ in Space, (Cambridge: Cambridge University Press),
        ed. F. Combes and G. Pineau des For\^ets, 13
\reference{} Savage, B. D., Drake, J. F., Budich, W., \& Bohlin, R. C.
	1977, \apj, 216, 291
\reference{} Savage, B. D. \& Sembach K. R. 1996, \araa, 34, 279
\reference{} Seidensticker, K. J. 1989, \aaps, 79, 61
\reference{} Seidensticker, K. J. \& Schmidt-Kaler, T. 1989, \aap, 225, 192
\reference{} Serkowski, K., Mathewson, D. L., \& Ford, V. L. 1975, \apj,
        196, 261
\reference{} Snow, T. P. 1993, \apj, 402, L73
\reference{} Snow, T. P., et al.\ 2000, ApJ, 538, L65
\reference{} Spitzer, L. \& Cochran, W. D. 1973, L23
\reference{} van Dishoeck, E. F. \& Black, J. H. 1986, \apj, 307, 332
\reference{} van Dishoeck, E. F., \& Black, J. H. 1989, \apj, 340, 273
\reference{} Whittet, D. C. B. \& van Breda, I. G. 1978, \aap, 66, 57
\reference{} Zhong, Z. P, Feng, R. F., Xu, S. L., Wu, S. L., Zhu, L. F.,
        Zhang, X. J., Ji, Q. \& Shi, Q. C. 1997, \pra, 55, 1799

\end{references}
\end{document}